\documentclass[a4paper]{article}
\usepackage[numbers,sort&compress]{natbib}
\usepackage{INTERSPEECH2020}

\title{Defense for Black-box Attacks on Anti-spoofing Models \\by Self-Supervised Learning}
\name{Haibin Wu$^1$, Andy T. Liu$^{1 2}$, Hung-yi Lee$^{1 2}$}

\address{
  $^1$Graduate Institute of Communication Engineering, National Taiwan University\\
  $^2$College of Electrical Engineering and Computer Science, National Taiwan University}
\email{\{f07921092, r07942089, hungyilee\}@ntu.edu.tw}

\begin{document}

\maketitle
\begin{abstract}
High-performance anti-spoofing models for automatic speaker verification (ASV), have been widely used to protect ASV by identifying and filtering spoofing audio that is deliberately generated by text-to-speech, voice conversion, audio replay, etc.
However, it has been shown that high-performance anti-spoofing models are vulnerable to adversarial attacks.
Adversarial attacks, that are indistinguishable from original data but result in the incorrect predictions, are dangerous for anti-spoofing models and not in dispute we should detect them at any cost.
To explore this issue, we proposed to employ Mockingjay, a self-supervised learning based model, to protect anti-spoofing models against adversarial attacks in the black-box scenario.
Self-supervised learning models are effective in improving downstream task performance like phone classification or ASR.
However, their effect in defense for adversarial attacks has not been explored yet.
In this work, we explore the robustness of self-supervised learned high-level representations by using them in the defense against adversarial attacks.
A layerwise noise to signal ratio (LNSR) is proposed to quantize and measure the effectiveness of deep models in countering adversarial noise.
Experimental results on the ASVspoof 2019 dataset demonstrate that high-level representations extracted by Mockingjay can prevent the transferability of adversarial examples, and successfully counter black-box attacks.

\end{abstract}


\section{Introduction}
\begin{figure*}[ht]
  \centering
  \centerline{\includegraphics[width=\linewidth]{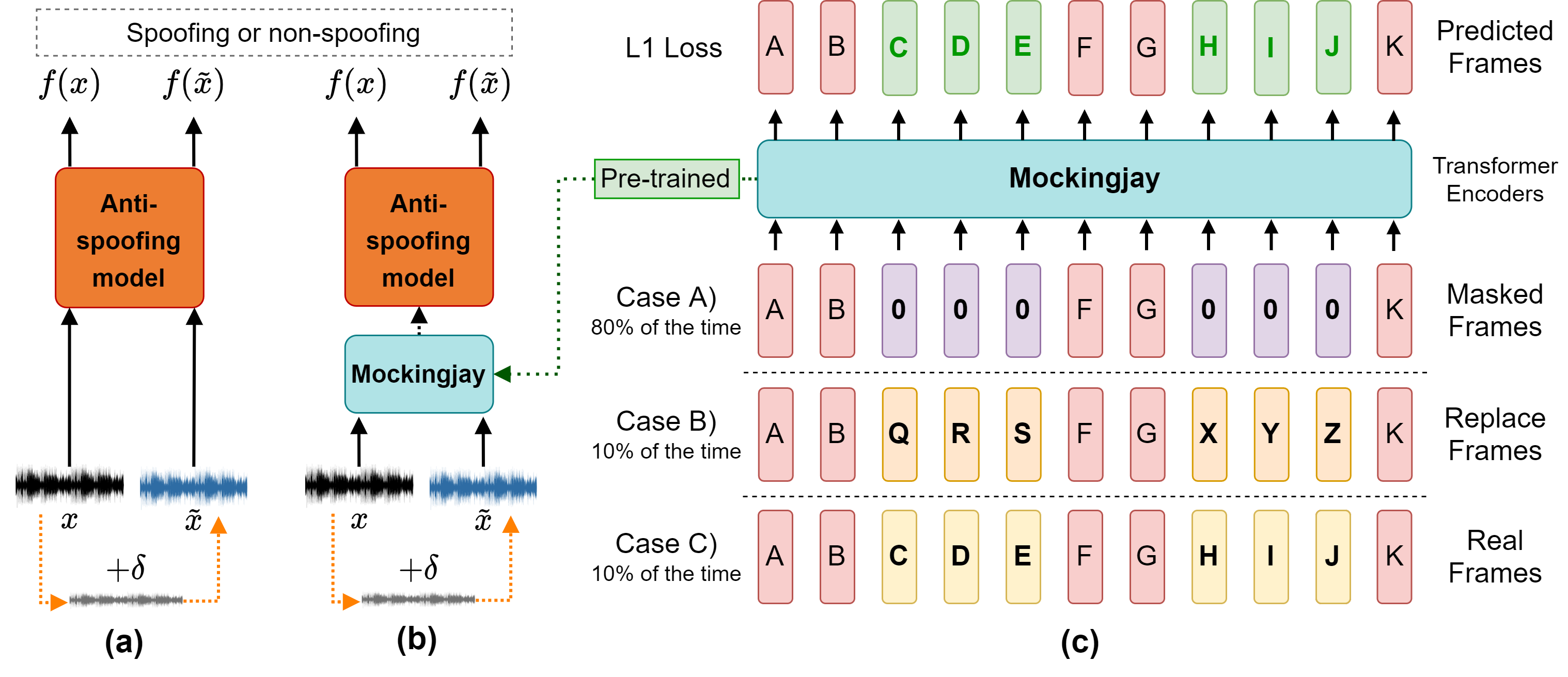}}
  \caption{(a) Adversarial attack, (b) Proposed method, (c) Illustration of the self-supervised Mockingjay pre-training.}
  \label{fig:framewrok}
\end{figure*}
Automatic speaker verification, abbreviated as ASV, is the task to verify whether a piece of speech sample belongs to a certain speaker. 
Thanks to the efforts of previous researchers \cite{snyder2018x,garcia2011analysis,reynolds2000speaker,lei2014novel,kenny2014deep}, it is now a matured technology widely applied to biometric identification. 
However, evidence shows that unprotected ASV models are highly vulnerable to spoofing audio deliberately generated by text-to-speech, voice conversion, and audio replay~\cite{wu2015spoofing,wu2016anti},
as some malicious attackers mimic a specific target user to deceive the ASV systems.
As a result, strategies to handle the spoofing audio are in need.
The ASVspoof challenge series \cite{wu2015asvspoof,kinnunen2017asvspoof,todisco2019asvspoof} is a community-driven challenge to arouse attention in addressing spoofing audio attacks and their countermeasures. 
Spoofing countermeasure models, also known as anti-spoofing models, are shields for ASV to detect and filter spoofing audio.
Recently several high performance anti-spoofing models have been proposed \cite{evans2013spoofing,Todisco_2016,delgado2018asvspoof,wu2012study,chen2015robust,chen2017resnet,qian2016deep,lai2019attentive,lavrentyeva2019stc,lai2019assert}.

Since Szegedy et al. \cite{szegedy2013intriguing} first proposed the concept of adversarial attacks, and illustrate how deep neural networks with impressive performance for computer vision tasks are vulnerable to adversarial attacks, a large variety of research in this domain have been done.
Adversarial example, which is generated by adding imperceptible perturbation to the input sample, can result in the incorrect prediction of the models.
The added perturbation is carefully crafted such that humans can not distinguish the adversarial example from the input sample visually or acoustically.
Attacking the models with adversarial examples is called adversarial attack.
Previous works show that deep neural networks for speech processing tasks are subject to adversarial attacks. 
\cite{carlini2018audio} investigates the vulnerability of automatic speech recognition (ASR) models to adversarial attacks.
Given any audio waveform, whether speech or music, they can generate an adversarial example, which is over 99\% similar to the original audio but makes the ASR model wrongly predict any transcribes they defined before. 
It has also been shown that ASV systems can be fooled by adversarial examples~\cite{kreuk2018fooling,das2020attacker, fli2020adversarial}.
Moreover, in \cite{wu2020defense,liu2019adversarial}, the authors illustrate the anti-spoofing models for ASV systems are also vulnerable to adversarial attacks.
In this paper, we focus on the defense for adversarial attacks of the anti-spoofing models. 

Proactive defense and passive defense are two main categories of defense for the adversarial attacks. 
The former defense aims to train new models to counter the adversarial attacks.
The most famous proactive defense method is adversarial training \cite{goodfellow2014explaining}, which injects the adversarial examples generated by different attack algorithms into the training data.
It is reasonable that the models are robust to specific attack algorithms if the models have already seen the adversarial examples during training.
However, adversarial training is time-consuming and resource-consuming.
What’s more, when defenders do adversarial training, they have no idea which attack algorithm the attackers will take. 
The mismatch between attack algorithms during training and inference will make the models susceptible to adversarial attacks they haven’t seen during training. 
Passive defense methods embrace the advantage of defending adversarial examples generated by all kinds of attacking methods without modifying the model. 
The spatial smoothing, also called as filter, is a passive defense method that counters adversarial attacks \cite{xu2017feature}. 
Gaussian filter, median filter and mean filter are used to defend anti-spoofing models in \cite{wu2020defense}. 
The above three filters are shallow filters to counter the deliberately generated adversarial perturbations. 

In this paper, we propose a passive defense leveraging the power of self-supervised learning. 
Self-supervised learning allows the model to learn high-level and contextualized representations from a large amount of unlabeled data, without the use of any label~\cite{bert}.
In self-supervised learning, a pre-training task (or auxiliary task) that uses only unlabeled data is formulated, and the model is required to solve such a task.
While solving the pre-training task, the model is also learning a function that maps input to high-level representations, which can potentially transfer information learned from unlabeled data to downstream tasks.
Through pre-training models on speech, self-supervised learning based models are able to leverage the knowledge of unlabeled speech, then the performance of downstream speech and language processing (SLP) tasks can be improved dramatically~\cite{mockingjay, cpc, apc, speech_encoder}, including phone classification, speaker recognition, and speech recognition.
However, the robustness of such high-level audio representations learned by self-supervised learning based models against adversarial attacks for anti-spoofing of ASV have not been studied yet.

In this work, we find that self-supervised learning models can serve as a deep filter which extracts the pivotal information from the contaminated input spectrograms to counter the adversarial attacks.
To the best of our knowledge, we are among the first ones to adopt the high-level representations extracted by the self-supervised model for the defense of anti-spoofing models in the black-box scenario and the experimental results show the effectiveness of our proposed method.
We also propose the layerwise noise to signal ratio (LNSR), to quantize and measure the effect of deep models in countering adversarial noise.
We find that the adversarial noise is attenuated layer by layer in the self-supervised learning model.


\section{Adversarial attack}


When a tiny perturbation, which is imperceptible to humans, is deliberately crafted and added to the original example, the new example will lead to the model's incorrect prediction.
We call the new example and the tiny perturbation as adversarial example and adversarial noise respectively.

As shown in Figure~\ref{fig:framewrok} a), $x$ is the original example, $\delta$ is the adversarial noise and $\Tilde{x}$ is the adversarial example.
Given the anti-spoofing model $f(.)$, we denote the prediction of the original example and adversarial example as $f(x)$ and $f(\Tilde{x})$.
The adversarial example is generated as this equation:
\begin{equation}
    \label{eq:1}
    \Tilde{x} = x + \delta. 
\end{equation}
Finding an adversarial example is equivalent to crafting a suitable perturbation $\delta$ and searching for a suitable $\delta$ is an optimization problem as shown below:
\vspace{-10pt}
\begin{equation}
    \label{eq:2}
    max_{\|\delta\|_{\infty} \leq \epsilon} Diff(f(x), f(\Tilde{x})),
\end{equation}
where $Diff(f(x), f(\Tilde{x}))$ means the difference between $f(x)$ and $f(\Tilde{x})$ and it is totally differentiable, $\|\delta\|_{\infty}$ is the $L_{\infty}$ norm, $\epsilon$ is a constant we defined to constrain $\delta$.
We solve the optimization problem by doing gradient descent to the input with the model parameters fixed.
We want the adversarial example $x$ to be as similar as $\Tilde{x}$ to make them indistinguishable by human, so the noise $\delta$ shouldn't be too large.
In this paper, $\delta$ is constrained in an $L_{\infty}$ norm ball.
Different searching strategies for $\delta$ result in different attack algorithms.
In this paper, we adopted the fast gradient-sign method (FGSM) \cite{goodfellow2014explaining} and the projected gradient descent method (PGD) \cite{madry2017towards}.


There are two kinds of adversarial attack scenarios: black-box attack and white-box attack. 
In both two scenarios, there are two models: the \textit{target} model and the \textit{attacking} model.
The target model is the model attackers aim to attack, and the attacking model is the model implemented by the attackers to generate adversarial examples.
In the black-box attack scenario, the target model and the attacking model are different models, while in the white-box attack scenario, the target model is also the attacking model.
In the white-box attack scenario, the attackers know everything about the target model, including model parameters, gradients, etc. 
Obviously it is unrealistic that the attackers have all access to the target model.
In the black-box scenario, the attackers can't obtain the inner parameters of the target model, while they can collect the inputs and the outputs of the target model by querying it. 
Then the attackers will train a substitute model and employ the substitute as the attacking model to generate adversarial samples with transferability.
Our objective in this paper is to prevent the transferability of such adversarial examples, and improve the robustness of anti-spoofing models against black-box attacks by leveraging high-level representations extracted by self-supervised models.

\section{Proposed method}

\subsection{Mockingjay}



The Mockingjay~\cite{mockingjay} approach learns representations of speech by solving a self-supervised masked-prediction task with a $L_1$ reconstruction loss function.
The model is based on multi-layer transformer encoders with multi-head self-attention~\cite{transformer} followed by a feed forward prediction network.
The transformer encoder produces a representation vector for each time frame, and the prediction network reconstructs frames of spectrogram to solve the masked-prediction task.
At training time, the masked-prediction task requires the model to take a sequence of frames as input that has had a certain percentage of randomly selected frames masked, and attempts to reconstruct the masked frames.
After training, the representations produced by the transformer network are inputs to the anti-spoofing model instead of acoustic features.
We illustrate this in Figure~\ref{fig:framewrok} b).

We consider a masking policy following \cite{mockingjay, bert}.
The following three cases are sampled. 
Case A) we mask all selected frames to zero; this happens for 80\% of the time. Case B) we replace all selected frames with random frames, 10\% of the time. And Case C) we leave all the selected frames untouched for the rest 10\% of the time.
Then the sampled case (A, B, or C) would be applied on 15\% of randomly selected frames.
The intuition is that by reconstructing from corrupted input, the model should learn a solid understanding of the high-level content, which provides immunity to adversarial attacks.
To adapt the local smoothness of acoustic sequence, we mask contiguous segments of $C_{num}$ of frames.
We illustrate this in Figure~\ref{fig:framewrok} c), where we show an example of $C_{num}=3$.
We also employ the downsampling technique from \cite{mockingjay}, where we reshape and stack $R_{factor}$ consecutive frames into one step.


\subsection{Self-Supervised Learned Adversarial Defender}
In this work, we propose to adopt the Mockingjay to protect the anti-spoofing models.
Superficial or surface features like Mel-Spectrogram often buries the abundant information of speech, extracting representations with the Mockingjay transform thus makes the high-level information more accessible to downstream tasks.
We first extract the high-level representations from spectrograms by Mockingjay and then use the high-level features to train the anti-spoofing model.
The cascade of the Mockingjay and anti-spoofing model shown in Figure~\ref{fig:framewrok} b) is called self-supervised learned adversarial defender.

In the black-box attack scenario, the attackers are not aware of the existence of the Mockingjay and only know the inputs to the target system are spectrograms.
They attempt to ﬁnd adversarial noise to add it to the input spectrograms by using the attacking model.
However, before the input spectrograms are thrown into the anti-spoofing model, the Mockingjay will help alleviate the superficial noise added to the input spectrograms and avoid the transferability of adversarial noise.
Experimental results show that the high-level representation extracted by Mockingjay prevents the transferability of adversarial noise and counter the black-box attacks. 

The readers may challenge the power of defense here comes from the mismatch of the network architecture between the target and attacking model. 
The target model has the Mockingjay as the front-end, while the attacking model does not.
It may be intuitive that the attack signal for the attacking model cannot transfer to the target model.
However, experimental results show that pre-training plays a critical role in the defense.
Without the pre-training, merely using the mismatch of network architecture can not avoid adversarial noise's transferability.

There are two plausible reasons that the Mockingjay can help counter the adversarial noise. 
From the perspective of the self-supervised training procedure of the Mockingjay, the masked-prediction task introduces noise to the input spectrograms.
The Mockingjay is trained to learn how to weaken the noise in the input spectrograms, extract pivotal information from the contaminated spectrograms, and use the pivotal information to reconstruct the original clean spectrograms.
The adversarial noise is also a kind of noise to some extent.
So in our proposed approach, the Mockingjay would weaken the adversarial noise added to the input spectrograms, extract key information and pass the key information to the anti-spoofing model to finish the anti-spoofing task.
From the loss function perspective, in the black-box attack scenario, usually the target model and the attacking model perform the same task and are trained by classification loss.
It is intuitively the adversarial perturbations generated by the attacking model are with transferability to the target model as they are both sensitive to classification loss.
However, in the proposed approach, the Mockingjay is trained by reconstruction loss and performs the task which is different from the attacking model.

\subsection{Layerwise noise to signal ratio} \label{subsec:LNSR}
We propose a measurement named layerwise noise to signal ratio ($LNSR$) in order to estimate the intensity of adversarial noise in different layers of the Mockingjay:
\begin{equation}
\begin{aligned}
    \label{eq:3}
    LNSR_{i} = \sum_{n=1}^{N}\frac{\|\hat{h}_{i}^{n}-h_{i}^{n}\|_{2}}{\|h_{i}^{n}\|_{2}}
    \\ for \, i=0,1, \ldots, K,
\vspace{-10pt}
\end{aligned}
\end{equation}
where $K$ is the total layer number of the Mockingjay, $N$ is the number of the adversarial example - original example pairs, $\|.\|_{2}$ means $L_{2}$ norm, $\hat{h}_{i}^{n}$ and $h_{i}^{n}$ are the features of the $i^{th}$ layer of the adversarial example and original example respectively for the $n^{th}$ pair. 
When $i=0$, $\hat{h}_{i}^{n}$ and $h_{i}^{n}$ are the adversarial example and original example themselves. 
$LNSR_{i}$ aims to measure the amount of the attack signals in each layer.
If the value of $LNSR_{i}$ decreases when $i$ increases, that means Mockingjay attenuates the attacking noises.

\begin{figure*}[ht]
  \centering
  \centerline{\includegraphics[width=\linewidth]{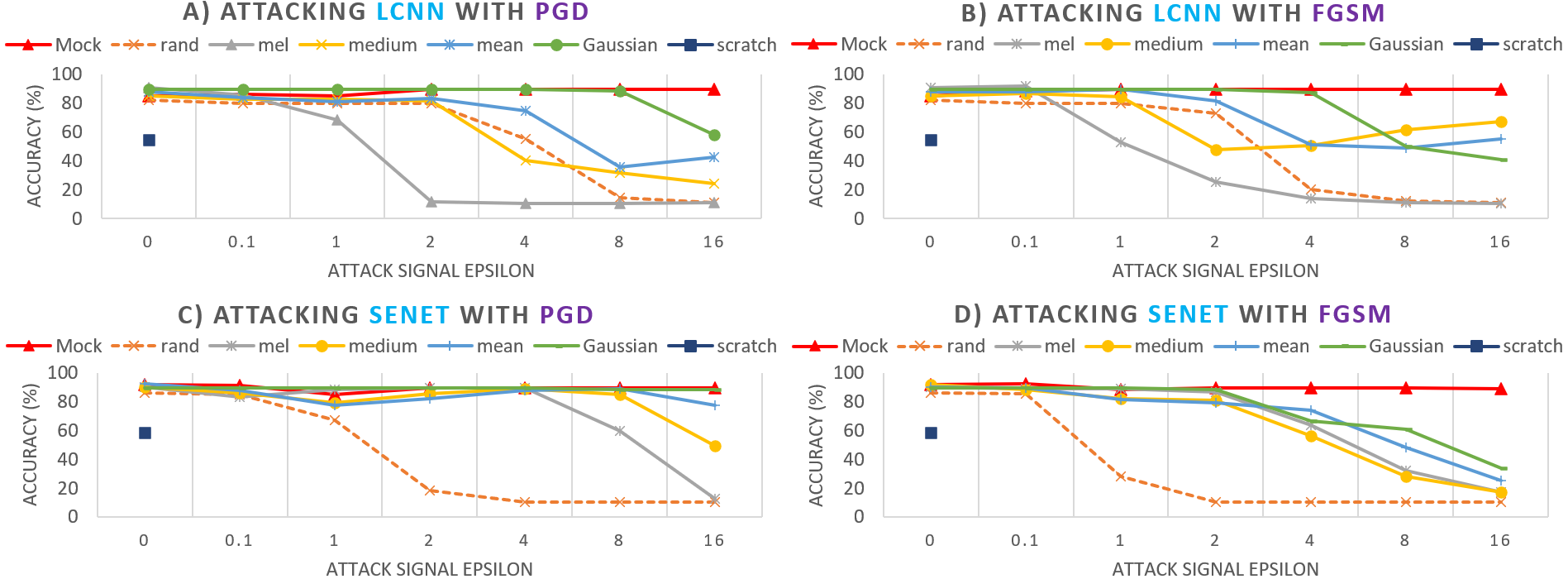}}
  \caption{Comparison of different defense methods against  two attack algorithms over increasing amount of attack signal $\epsilon$.}
  \label{fig:result}
\end{figure*}

\begin{figure}[h]
  \centering
  \centerline{\includegraphics[width=\linewidth]{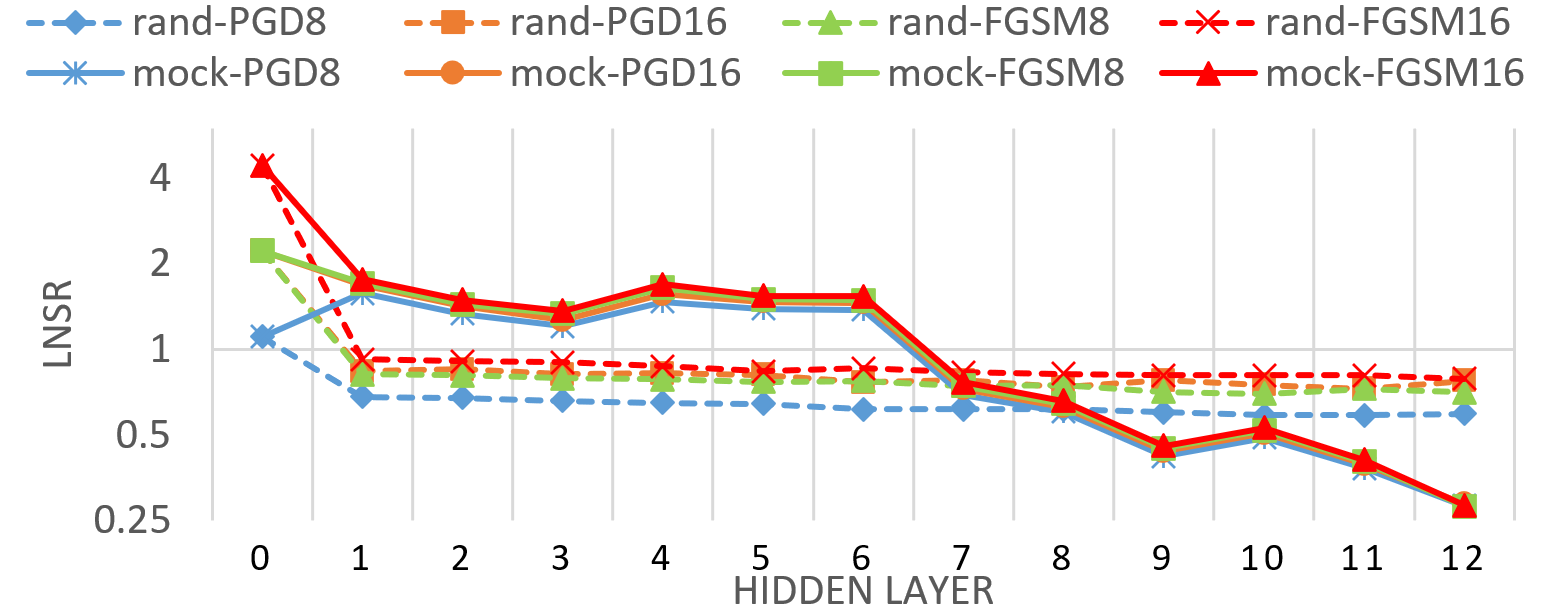}}
  \caption{Layerwise noise to signal ratio on Mockingjay}
  \label{fig:lnsr}
  \vspace{-10pt}
\end{figure}

\section{Experiment}
\subsection{Experiment setup}
For the dataset, we use the LA partition of the ASVspoof 2019 challenge, which contains fake audios generated by text to speech and voice conversion.
The dataset is itself divided into three parts: training, development, and evaluation.

For the Mockingjay, we use the prevailing framework of the LARGE model described in Mockingjay~\cite{mockingjay}, which consists of 12 layers of Transformer Encoders.
We follow the pre-training settings as in \cite{mockingjay}, where we pre-train our Mockingjay model on 360 hours of speech on the LibriSpeech dataset~\cite{librispeech}.
Two high-performance anti-spoofing models are adopted: LCNN \cite{lavrentyeva2019stc} and SENet \cite{lai2019assert}. The implementation details of the two models can be found in \cite{liu2019adversarial}.
We refer to the LCNN and SENet trained by the mel-spectrograms as basic LCNN and SENet.
In the black-box attack scenario, we use the basic LCNN and SENet as the attacking models to generate adversarial examples and use the models in Figure~\ref{fig:result} as target models. 
The adversarial examples generated by the basic LCNN are used to attack the basic SENet and its variants. 
The adversarial examples generated by the basic SENet are used to attack the basic LCNN and its variants.
We use FGSM and PGD as attack algorithms, and we measure over different values of $\epsilon$: $0.1, 1, 2, 4, 8, 16$.

\subsection{Result and analysis}


\subsubsection{Comparing different defense approaches}

The proposed approach is compared with various passive filter-based defense approaches~\cite{wu2020defense}.
Results are presented in Figure~\ref{fig:result}. 
We refer to the cascade of the pre-trained Mockingjay and anti-spoofing model as Mock,
the basic LCNN or SENet as Mel, and finally the anti-spoofing models equipped with different hand-designed filters as medium, mean, and Gaussian.
In Figure~\ref{fig:result} A) and B), LCNN and its variants are attacked by PGD and FGSM, respectively.
In Figure~\ref{fig:result} C) and D), SENet and its variants are attacked by PGD and FGSM, respectively.

As expected, basic Mel models (grey curve) are vulnerable to adversarial attacks.
In all four scenarios, we see the proposed Mockingjay defense mechanism prevails over all other approaches (red curve, denoted as Mock), as Mockingjay is invariant to adversarial attacks.
The attack always fails no matter the amount of attack signal.
Although Mockingjay is pre-trained on LibriSpeech~\cite{librispeech} but not ASVspoof data, it is still capable of leveraging self-supervised learned knowledge to defense adversarial attacks.
Other filters (medium, mean, and Gaussian) also counter the attack to some extent, but cannot resist high values of $\epsilon$ as Mockingjay.
The Mockingjay model outperforms all the filters in all circumstances.

Moreover, we show results of a random parameterized Mockingjay (orange curve, denoted as rand) to demonstrate the effect of pre-training.
Random parameterized Mockingjay also shows some capability of defense in Figure~\ref{fig:result} A) and B), but fails to protect anti-spoofing models against adversarial attacks as $\epsilon$ increases. 
It completely fails to protect anti-spoofing models in Figure~\ref{fig:result} C) and D).
The results show that the random Mockingjay is much worse than the pre-trained model, and even worse than the hand-designed filters in some cases.
This shows that our success in the red curve (Mock) is not simply from the mismatch of network architecture between the target model and attacking model.
To further show the importance of pre-training, we trained the anti-spoofing models with the same architecture as the cascade of Mockingjay and LCNN/SENet from scratch. 
The results are denoted as ``scratch'' (dark blue dot).
The results show that training the cascade model from scratch results in a low accuracy that barely surpasses random guesses.
This further shows that the success in the red curve (Mock) is not contributed by model size.

\subsubsection{Measuring the removal of adversarial noise}
The values of $LNSR$ (Section~\ref{subsec:LNSR}) in different layers of different models are shown in Figure~\ref{fig:lnsr}.
rand-PGD$\epsilon$ and mock-PGD$\epsilon$ means we use the adversarial examples generated by PGD with $\epsilon$ to calculate the $LNSR$ on the random parameterized Mockingjay and pre-trained Mockingjay, respectively. 
Two $\epsilon$ values are tested: $8,16$.
From Figure~\ref{fig:lnsr}, the pre-trained Mockingjay successfully lowers the $LNSR$.
When the model becomes deeper, the value of $LNSR$ becomes lower, which illustrates the effect of the Mockingjay to alleviate the adversarial noise.
In contrast, the random parameterized Mockingjay can only reduce $LNSR$ to a certain degree. 
When model depth increases, the value of $LNSR$ is quickly saturated.
This is another evidence to show the importance of pre-training in defense.

\section{Conclusion}
In this work, we propose to use a self-supervised learning model to protect the anti-spoofing models against black-box attacks. 
Experimental results illustrate the representations extracted by self-supervised learning model prevent the transferability of adversarial examples and counter the black-box attacks.
The proposed layerwise noise to signal ratio manifests the effectiveness of the self-supervised learning model in alleviating the adversarial noise layer by layer.
For the future work, we would like to explore the capability of defense for the self-supervised model learned with different objectives and apply the proposed defense approaches on more speech processing applications.


\bibliographystyle{IEEEtran}

\bibliography{reference.bib}

\end{document}